\documentclass[fleqn,usenatbib]{mnras}

\usepackage{newtxtext,newtxmath}
\usepackage[T1]{fontenc}
\usepackage{ae,aecompl}

\usepackage{graphicx}	% Including figure files
\usepackage{amsmath}	% Advanced maths commands
\usepackage{amssymb}	% Extra maths symbols
\usepackage{booktabs}
\usepackage{multirow}
\usepackage{siunitx}
\usepackage{comment}
\usepackage{placeins}
\usepackage[shortlabels]{enumitem}
\newcommand{\gpe}{GP-EBOP}

\newcommand{\Nstar}{NGTS-2}
\newcommand{\Nplanet}{NGTS-2b}
\newcommand{\Nfield}{NG1416-3056}

\newcommand{\vmag}{$10.96$} % APASS
\newcommand{\VSini}{$15.2\pm0.8$} % Barry
\newcommand{\Prot}{$5.66^{+0.49}_{-0.44}$} % Barry's vsini, i=90
\newcommand{\age}{$2.17\pm0.37$} % SPECIES
\newcommand{\NstarTeff}{$6478^{+94}_{-89}$} % SED final
\newcommand{\FeH}{$-0.06\pm0.09$} % Barry
\newcommand{\NstarRad}{$1.702 ^{+0.047}_{-0.044}$} % SED final
\newcommand{\Nstarlogg}{$4.197^{+0.030}_{-0.059}$} % global modelling
\newcommand{\Nstardense}{$0.477^{+0.030}_{-0.061}$} % global modelling
\newcommand{\NstarMass}{$1.64^{+0.19}_{-0.22}$} % Msun, global modelling

\newcommand{\NperiodShort}{$4.51$} % days
\newcommand{\NplanetDepth}{$1.0\%$} % visual inspection
\newcommand{\NplanetDuration}{$4.651^{+0.046}_{-0.038}$} % hours
\newcommand{\Ngamma}{$-26.3616^{+0.0064}_{-0.0063}$} % km/s
\newcommand{\NplanetRad}{$1.595^{+0.047}_{-0.045}$} % Rjup
\newcommand{\NplanetMass}{$0.74^{+0.13}_{-0.12}$} %Mjup
\newcommand{\NplanetDense}{$0.226^{+0.040}_{-0.038}$} % gcm^-3
\newcommand{\NplanetTeff}{$1468^{+45}_{-42}$} % K
\newcommand{\SemiMajorAxis}{$0.0630^{+0.0024}_{-0.0030}$} % AU
\newcommand{\OrbitalInc}{$88.5^{+1.0}_{-1.2}$} % deg

\newcommand{\NstarPMRA}{$-21.736\pm0.093$} % Gaia
\newcommand{\NstarPMDec}{$-0.858\pm0.090$} % Gaia
\newcommand{\GalULSR}{$30.048 \pm 0.028$} % SH
\newcommand{\GalVLSR}{$25.781 \pm 0.020$} % SH
\newcommand{\GalWLSR}{$-5.093 \pm 0.018$} % SH
\newcommand{\Ndist}{$360.3 ^{+8.3}_{-7.8}$} % SED final

\newcommand{\kms}{km\,s$^{-1}$}
\newcommand{\kmssquared}{km$^2\,s^{-2}$}
\newcommand{\ms}{m\,s$^{-1}$}

\newcommand{\luminosity}{erg\,s$^{-1}$}
\newcommand{\cmss}{cm\,s$^{-2}$}

\newcommand{\gcmthree}{g\,cm$^{-3}$}
\newcommand{\masy}{mas\,yr$^{-1}$}
\newcommand{\mjup}{M$_{J}$}
\newcommand{\rjup}{R$_{J}$}

\newcommand{\msun}{$M_{\odot}$}

\newcommand{\teff}{$T_{\rm eff}$}
\newcommand{\arcsecpix}{arcsec\,pixel$^{-1}$}

\newcommand{\LSO}{La Silla Observatory}

\newcommand{\JWST}{{\it JWST}}
\newcommand{\NGTS}{{\it NGTS}}

\newcommand{\TESS}{{\it TESS}}

\renewcommand{\footnote}{\textsuperscript{\arabic{footnote}}}

\setlist[enumerate]{%
labelsep=6pt,%                           
labelindent=0.\parindent,%               
itemindent=0pt,%
leftmargin=*,%                          
listparindent=-\leftmargin% 
}

\title[\Nplanet: An inflated hot-Jupiter]{\Nplanet: An inflated hot-Jupiter transiting a bright F-dwarf}

\author[Liam Raynard]{
\parbox{\textwidth}{
Liam Raynard,$^{l}$\thanks{E-mail: \href{lr182@le.ac.uk}{lr182@le.ac.uk}}
Michael R. Goad,$^{l}$
Edward Gillen,$^{c,\dagger}$
Louise D. Nielsen,$^{g}$
Christopher~A.~Watson,$^{q}$
Andrew~P.~G.~Thompson,$^{q}$
James McCormac,$^{w, ce}$
Daniel~Bayliss,$^{w,ce}$
Maritza~Soto,$^{uc}$
Szilard~Csizmadia,$^{d}$
Alexander Chaushev,$^{l}$
Matthew R. Burleigh,$^{l}$
Richard Alexander,$^{l}$
David J. Armstrong,$^{w,ce}$
Fran\c{c}ois Bouchy,$^{g}$
Joshua T. Briegal,$^{c}$
Juan~Cabrera,$^{d}$
Sarah L. Casewell,$^{l}$
Bruno Chazelas,$^{g}$
Benjamin~F.~Cooke,$^{w, ce}$
Philipp~Eigm\"uller,$^{d,tu}$
Anders~Erikson,$^{d}$
Boris~T.~G\"ansicke,$^{w,ce}$
Andrew Grange,$^{l}$
Maximilian~N.~G{\"u}nther,$^{c}$
Simon~T.~Hodgkin,$^{ca}$
Matthew J. Hooton$,^{q}$
James~S. Jenkins,$^{uc,ci}$
Gregory Lambert,$^{c}$
Tom Louden,$^{w,ce}$
Lionel Metrailler,$^{g}$
Maximiliano~Moyano,$^{a}$
Don~Pollacco,$^{w,ce}$
Katja Poppenhaeger,$^{q}$
Didier~Queloz,$^{c,g}$
Roberto~Raddi,$^{rs,w}$
Heike~Rauer,$^{d,tu,fu}$
Andrew M. Read,$^{l}$
Barry~Smalley,$^{k}$
Alexis~M.~S.~Smith,$^{d}$
Oliver Turner,$^{g}$
St\'{e}phane~Udry,$^{g}$
Simon.~R.~Walker,$^{w}$
Richard~G.~West,$^{w,ce}$
Peter~J.~Wheatley$^{w,ce}$
}
\\
\\Affiliations are listed at the end of the paper.
}

\date{Accepted for publication on 18 September 2018}

\pubyear{2018}

\begin{document}
\label{firstpage}
\pagerange{\pageref{firstpage}--\pageref{lastpage}}
\maketitle

% Abstract of the paper
\begin{abstract}
We report the discovery of \Nplanet, an inflated hot-Jupiter transiting a bright F5V star (2MASS J14202949-3112074; \teff=\NstarTeff~K),
discovered as part of the Next Generation Transit Survey (\NGTS). The planet is in a P=\NperiodShort~day orbit with mass \NplanetMass~\mjup{}, radius \NplanetRad~\rjup{} and density \NplanetDense{}~\gcmthree; therefore one of the lowest density exoplanets currently known. With a relatively deep \NplanetDepth{} transit around a bright V=\vmag\ host star, \Nplanet{} is a prime target for probing giant planet composition via atmospheric transmission spectroscopy. The rapid rotation ({\it v}\,sin\,{\it i}=\VSini{} \kms) also makes this system an excellent candidate for Rossiter-McLaughlin follow-up observations, to measure the sky-projected stellar obliquity. \Nplanet{} was confirmed without the need for follow-up photometry, due to the high precision of the \NGTS{} photometry.
\end{abstract}

% Select between one and six entries from the list of approved keywords.
% Don't make up new ones.
\begin{keywords}
techniques -- photometric -- stars:individual: NGTS-2, planetary systems
\end{keywords}

%%%%%%%%%%%%%%%%%%%%%%%%%%%%%%%%%%%%%%%%%%%%%%%%%%

%%%%%%%%%%%%%%%%% BODY OF PAPER %%%%%%%%%%%%%%%%%%

\section{Introduction}
\label{sec:intro}
Hot-Jupiters are giant exoplanets $(M\gtrsim 0.5$~\mjup{}) orbiting close to their parent stars ($P\lesssim 10$ days). Due to the increased stellar irradiation they experience, hot-Jupiters (hereafter HJs) have higher effective temperatures and larger radii ($1\lesssim R \lesssim 2~$\rjup{}) compared to cooler gas giants at larger orbital distances, such as Jupiter \citep{Schneider2011,Laughlin2011,Santerne2016}. However, even when accounting for increased stellar irradiation, the radii of many HJs exceed that predicted by evolutionary models \citep{Baraffe2003,Burrows2007} and HJs with bulk densities as low as $\sim0.1$ \gcmthree{} have been discovered \citep{Smalley2012,Hartman2016}. Various internal heating mechanisms have been proposed to reconcile the problem of inflated HJ radii but a proper understanding among the community is still developing \citep{Baraffe2010,Fortney2010,Baraffe2014,Thorngren2018,Sestovic2018}.

HJs transiting bright stars present the finest opportunities for robust exoplanet atmospheric characterisation.  The twelve community targets for the James Webb Space Telescope (\JWST; \citealt{Gardner2006,Kalirai2018}) presented in \citet{Stevenson2016} are all gas giants transiting host stars brighter than $V=12$ with orbital periods $<5$\,days.  It is therefore important to discover and accurately characterise such systems in advance of \JWST\ operations.  Already, comparative atmospheric transmission studies have revealed a diverse range of HJ atmospheres, ranging from clear to cloudy \citep{sing2016}.  Only by discovering new HJs transiting bright stars will we be able to expand these studies and look for statistically significant correlations that may shed light on the formation and evolution of these planets and their atmospheres.

The majority of early exoplanet discoveries were HJs, and although we now recognise that HJs are relatively rare \citep[found around only $\simeq0.5$\% of Sun-like stars][]{Fressin2013}, they still represent important benchmarks for planet formation theories. HJs almost certainly formed at larger orbital separations and migrated to their observed, short-period orbits, but the migration mechanism remains uncertain. Some HJs have orbits that are eccentric and/or misaligned to the stellar rotation axis \citep[see][]{triaud10,jenkins17}, and therefore inconsistent with the predictions of disc-driven migration, but the incidence rate of HJs is similarly difficult to reconcile with tidal (``high eccentricity") migration \citep[see the review by][and references therein]{dawson18}. Further detections are therefore crucial if we are to understand how HJs form and evolve.

In this paper we report the discovery of \Nplanet, a HJ transiting a bright (V=\vmag) star in a \NperiodShort\,day orbit.  In Section \ref{sec:observations} we present the \NGTS\ photometric observations that led to the discovery of this planet, as well as the HARPS spectroscopic observations that confirmed the planet and allowed us to determine its mass.  In Section \ref{sec:analysis} we analyse the available data to characterise the system and determine its properties. Finally, in Section \ref{sec:discussion} we present our conclusions and discuss our results in the context of similar known systems and the opportunities for further follow-up with upcoming space missions.

\section{Observations}
\label{sec:observations}
\Nplanet{} was discovered using the \NGTS{} telescopes in conjunction with high precision spectroscopy from HARPS.  We detail  these observations in this Section.  

\subsection{NGTS Photometry}
\label{sec:photom}
\NGTS{} is a wide-field, transit survey located at ESO's Paranal Observatory in Chile. The primary goal of \NGTS{} is to discover Neptune sized exoplanets orbiting bright (V<13), K and early M-dwarfs -- suitable for atmospheric follow-up studies. \NGTS{} comprises an array of twelve fully automated, 20\,cm Newtonian telescopes, mounted to independent equatorial forks. Each telescope is coupled to an Andor Ikon-L Camera featuring a 2K$\times$2K e2V deep-depleted, red-sensitive CCD, with 13.5\,$\mu$m pixels and an instantaneous field of view of 8 deg$^2$. Further details on \NGTS{} can be found in \citet{Wheatley2018}.

Field \Nfield{} was observed by a single \NGTS{} telescope and camera in closed-loop autoguiding mode \citep{McCormac2013}, over a photometric campaign conducted between 2017 January 02 and 2017 August 21.  In total 199,324$\, \times$10\,s exposures were obtained in the \NGTS{} bandpass (520 -- 890\,nm) over 139 usable nights. The tracking of the \Nfield{} field over the 139 nights was stable to an RMS of $0.28$ and $0.11$ pixels in the X and Y directions, respectively.

Raw data were processed by the \NGTS{} pipelines, to obtain systematic detrended light curves for our target object catalogue. A full description of the \NGTS{} pipelines can be found in \citet{Wheatley2018}.

\subsubsection{Planet detection and vetting}
\label{subsubsec:planet_detection_screening}
\begin{figure*}
	\centering
    \includegraphics[width=.7\paperwidth]{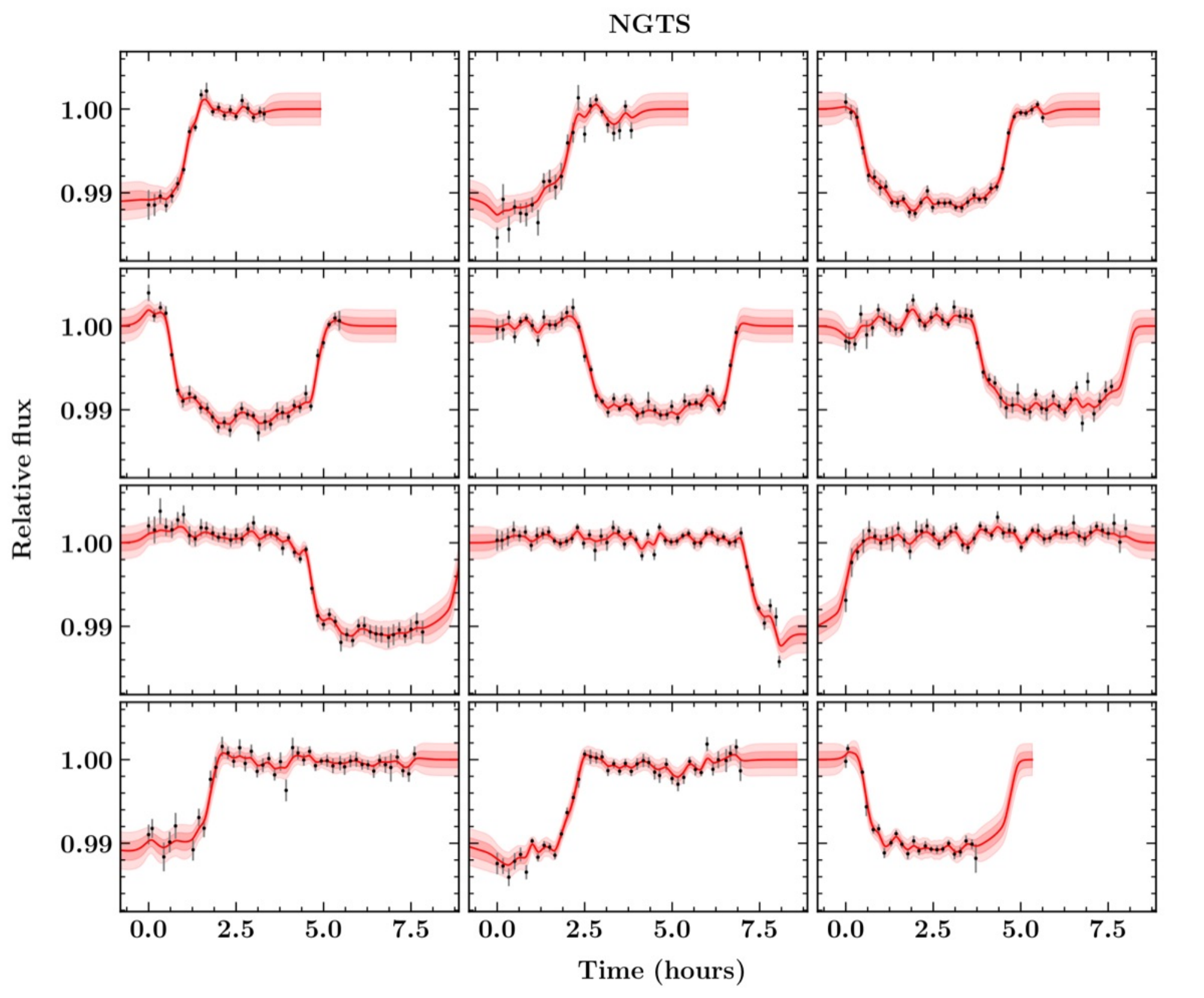}
    \caption{Individual transits of \Nstar{} detected in the \NGTS{} light curve. Black points represent photometric data binned to 10-minute cadence. The red line and pink shaded regions show the median and 1 \& 2\,$\sigma$ confidence intervals of the posterior model using GP-EBOP \citep{gillen17} set out in Section~\ref{subsec:global_modelling}, before detrending for the Gaussian process component. The robust detection of \Nplanet{} in individual transits (RMS$\sim2.4$ mmag at 10 minute sampling) demonstrates the high photometric precision of \NGTS{}.}
    \label{fig:NGTS_individual_transits}
\end{figure*}

\begin{figure}
    \includegraphics[width=\columnwidth]{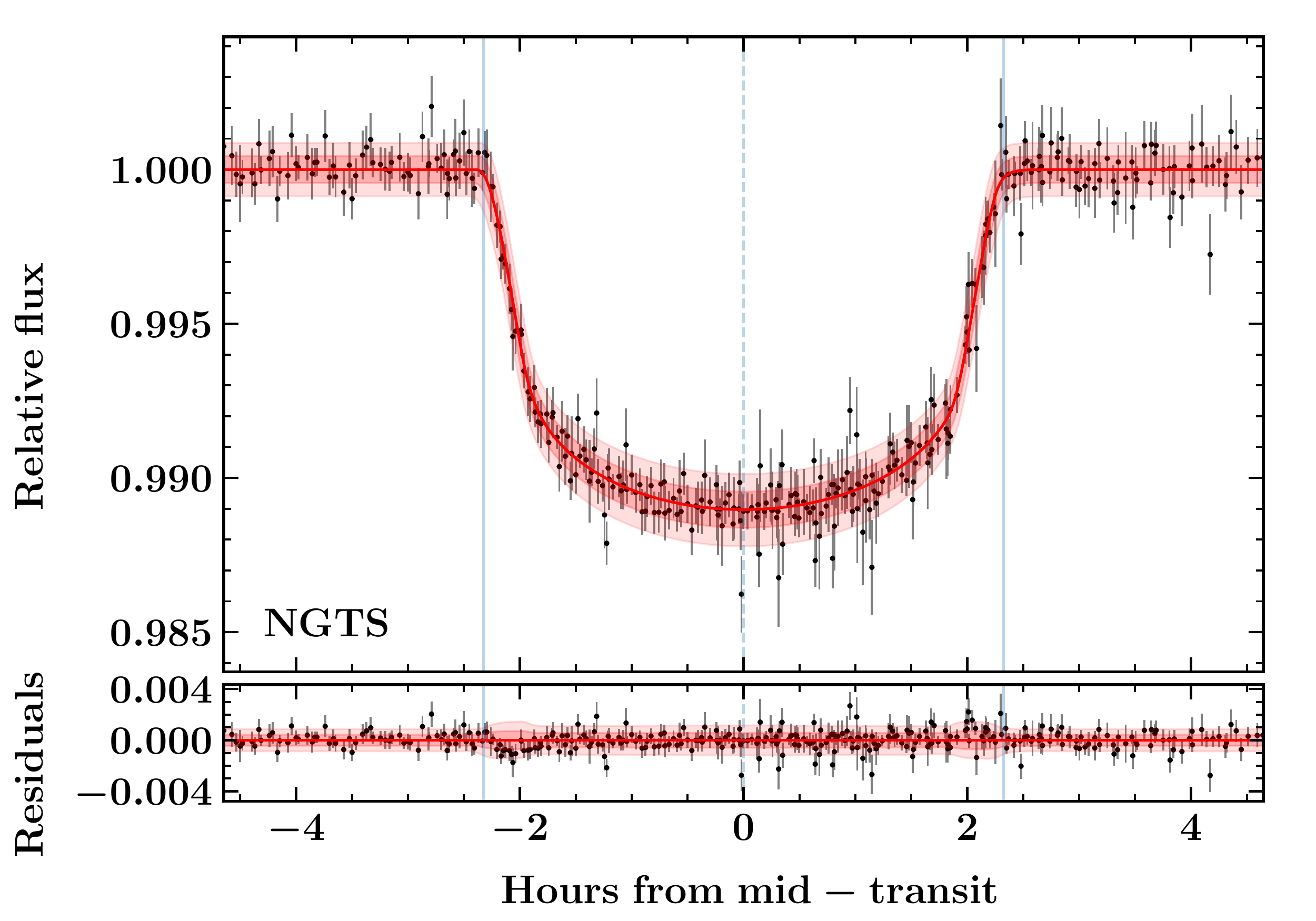}
    \caption{Transit of \Nstar{}, phase-folded on the best fitting period as determined from the global modelling set out in Section~\ref{subsec:global_modelling}. Black points represent photometric data binned to 10-minute cadence. The red line and pink shaded regions show the median and 1 \& 2\,$\sigma$ confidence intervals of the posterior model using GP-EBOP, which has been detrended for the Gaussian process component. The vertical blue lines represent the transit centre and first and fourth contact points. Residuals are shown below with \mbox{RMS $\sim1.3$ mmag.}}
    \label{fig:NGTS_phase_zoomed}
\end{figure}

We searched the $\sim$11,000 object light curves from the \Nfield{} field for transit-like signals using ORION \citep{Wheatley2018} - an implementation of the box-fitting least squares (BLS) algorithm \citep{Kovacs2002}. ORION identified a strong \NplanetDepth{} depth transit signal for \Nstar{} with a period of $\sim$4.5 days, derived from 12 individual transits. We present the individual transit light curves in Figure \ref{fig:NGTS_individual_transits} and the full light curve, phase-folded to the best fitting period as determined from our global modelling analysis (Section \ref{subsec:global_modelling}), in Figure \ref{fig:NGTS_phase_zoomed}. A portion of the full light curve dataset is provided in Table~\ref{tab:ngts}.

In order to screen for false-positives mimicking a planetary transit signal, we applied a series of vetting tests. First, we see no evidence of a secondary eclipse at phase 0.5, which would have indicated that the signal is due to an eclipsing stellar companion. Second, we see no depth difference in odd/even numbered transits, which tells us we have identified the true period as opposed to half the true period.

Out of transit ellipsoidal variations are commonly observed for short-period stellar binaries. We do see evidence of sinusoidal variation with a period of $\sim$10.8 hours, which coincides with boundaries between regions of differing data point density. Therefore we attributed the variability to an observation window effect, as opposed to evidence for a stellar binary, a conclusion supported by a low radial velocity amplitude (see Section \ref{sec:spect}).

\Nstar{} has a relatively small proper motion of \mbox{PM$_{\mathrm{RA}}$=\NstarPMRA~and PM$_{\mathrm{Dec}}$=\NstarPMDec~\masy} \citep{Gaia2018}. Analysis of Gaia sources shows no other objects within the 0.4 arcsec limit of Gaia DR2. Rain diagrams were famously generated for Kepler candidates to identify correlations between photometric flux and centroid time series \citep{Batalha2010,Bryson2013}. In a similar way, we were able to look for correlations between transit events and shifts in the centre of photometric flux using the method from \citet{Guenther17b}. This technique is able to detect false-positive signals due to background contaminating objects within $\sim$1\arcsec, much smaller than the size of individual \NGTS{} pixels ($\sim$5\arcsec). We find no centroiding correlations down to the milli-pixel level in this case.

Finally, we utilised the astrometric and photometric parameters in Table \ref{tab:stellar}, in conjunction with stellar SED modelling and stellar populations from the Besan\c{c}on Galaxy model \citep{robin03}, to determine that \Nstar{} is an F-dwarf (F5V) rather than an F-giant. This assured us that the observed (\NplanetDepth{}) transit depth can be caused by an occulting body of planetary radius.

We searched for the existence of additional transiting planets around \Nstar{} by masking the transit of \Nplanet{} in the \NGTS{} light curve and conducting a series of 5 additional BLS runs. At each step we masked and removed any `in-transit' data points before feeding the remaining data into the next iteration. We searched the period range $0.425$ to $30.0$ days, with a period step of 1\,min. We find no other significant BLS detections that resemble a transiting planet in the \NGTS{} light curve. 

In conclusion, the planet candidate \Nplanet{} passed all of our screening tests and we therefore scheduled the target for spectroscopic follow-up. We note that, in contrast to typical ground-based exoplanet discoveries, follow-up photometry was unnecessary when vetting \Nplanet{} owing to the photometric precision of the \NGTS{} light curve (RMS$\sim2.4$ mmag) in conjunction with the overall rigorous vetting process applied.

\begin{table}
	\centering
	\caption{\NGTS{} photometry for \Nstar.  The full table is available in a machine-readable format from the online journal.}
	\label{tab:ngts}
	\begin{tabular}{ccc}
	BJD$_\mathrm{TDB}$ (-2,450,000) &	Flux (normalised)       	&Flux error\\
	\hline
	7755.83964926 & 1.001100 & 0.001087 \\
	7755.84844136 & 0.998410 & 0.001036 \\
	7755.85719473 & 0.998299 & 0.0008004 \\
	7756.83718233 & 0.998381 & 0.001419 \\
	7756.84592365 & 1.003388 & 0.0009239 \\
	7756.85470943 & 0.999316 & 0.001139 \\
	7756.86076224 & 1.002791 & 0.001479 \\
	7757.83477136 & 0.998151 & 0.001009 \\
	7757.84352087 & 0.997511 & 0.0007974 \\
	7757.85230863 & 0.998330 & 0.0007950 \\
        ...        &   ...    &   ...   \\
	\hline
	\end{tabular}
\end{table}

\subsection{Spectroscopy and radial velocities}
\label{sec:spect}
Spectroscopic data of \Nstar{} were acquired with the HARPS spectrograph \citep{Mayor2003} on the ESO 3.6\,m telescope at \LSO, Chile under programme 099.C-0303(A) and 0100.C-0474(A). A total of 10 measurements, in High Accuracy Mode (hereafter HAM), were taken between 2017 July 25 and 2017 September 5. An additional 6 measurements were taken in EGGS (high efficiency) mode between 2017 August 17 and 2018 March 12. 

Radial velocities (RV) were calculated via cross-correlation with a G2 binary mask, using the standard HARPS reduction pipeline. One measurement was corrected for moonlight contamination, by subtracting the flux from the simultaneous sky fibre. Initial analysis of the RV data showed a variation in phase, consistent with the orbital period and epoch derived from the \NGTS{} photometric data. The semi-amplitude was K$\sim65$\,\ms, from which we inferred the existence of a planetary companion. 

\begin{figure}
\includegraphics[width=\columnwidth]{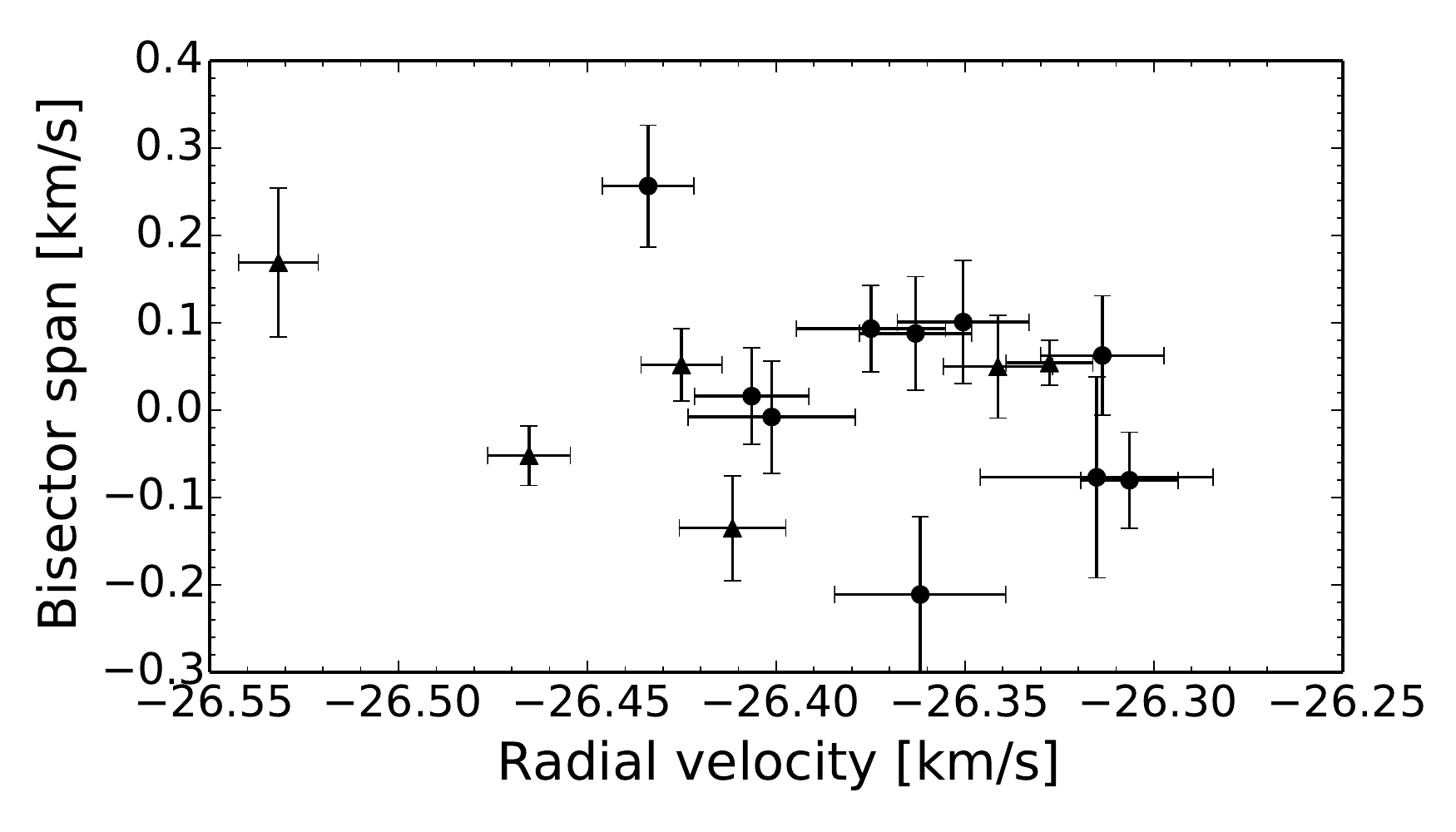}
\includegraphics[width=0.97\columnwidth]{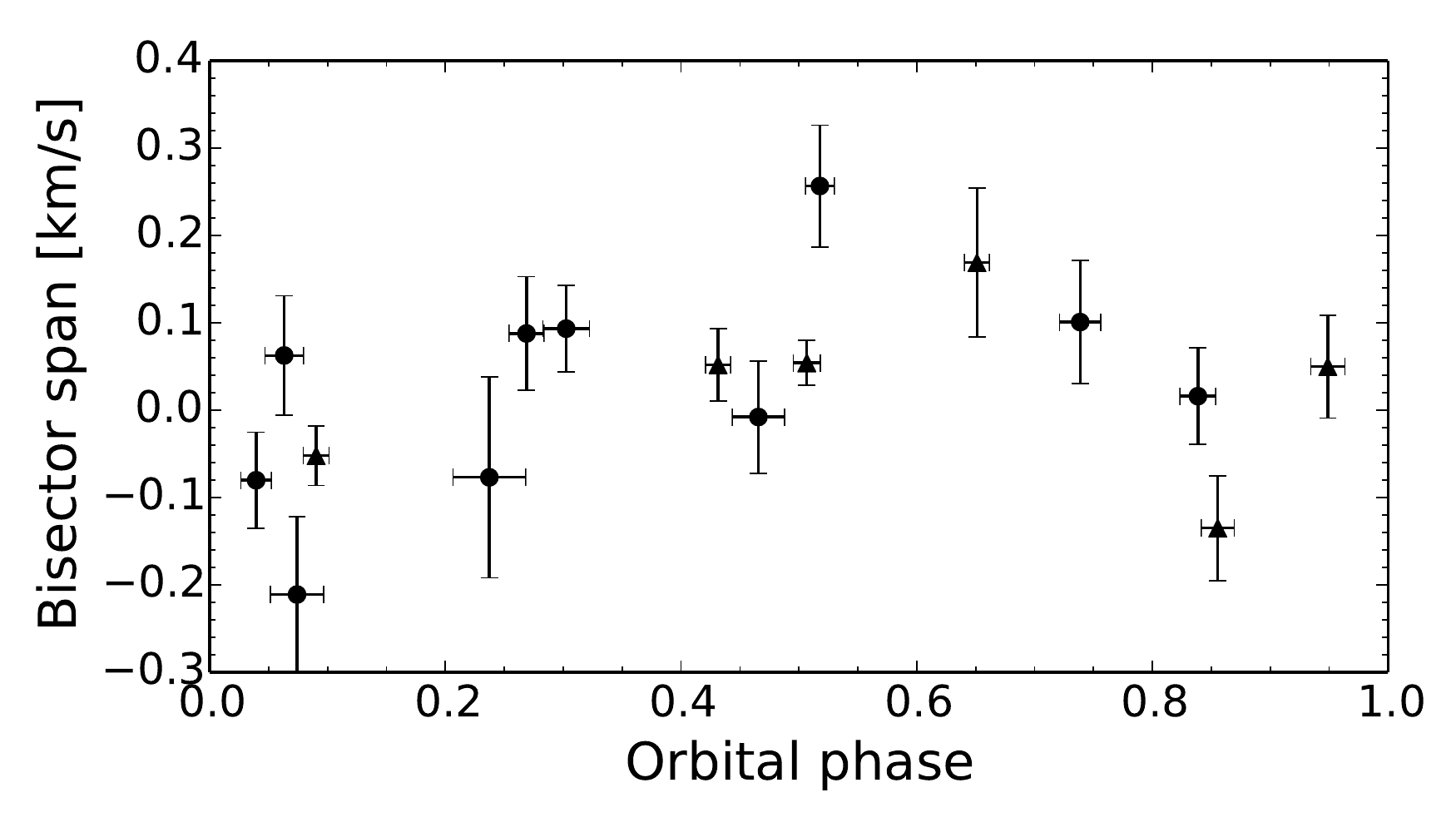}
    \caption{Line bisector span for each HARPS radial velocity measurement, plotted against the measured radial velocity (top panel) and orbital phase (bottom panel). Black points represent HARPS/HAM data points whereas black triangles show HARPS/EGGS mode data. Despite the large variation in bisector span, we find no correlation with the radial velocities.}
    \label{fig:bis}
\end{figure}

To ensure that the RV signal does not originate from cool stellar spots or a blended eclipsing binary, we checked for correlations between the line bisector spans and the RV measurements (Figure \ref{fig:bis}). The bisector span was calculated using a modified version of the standard approach from \citet{Queloz2001}, where we disregarded the bottom 20~\% of the peak of the cross-correlation function (CCF), while averaging over a greater span. This gives a robust measurement of the asymmetry of the CCF peak, which is less sensitive to local effects caused by stellar pulsation. We find no evidence for a correlation but note a large variation in bisector, suggesting that \Nstar{} is an active or pulsating star.

\begin{figure}
\includegraphics[width=\columnwidth]{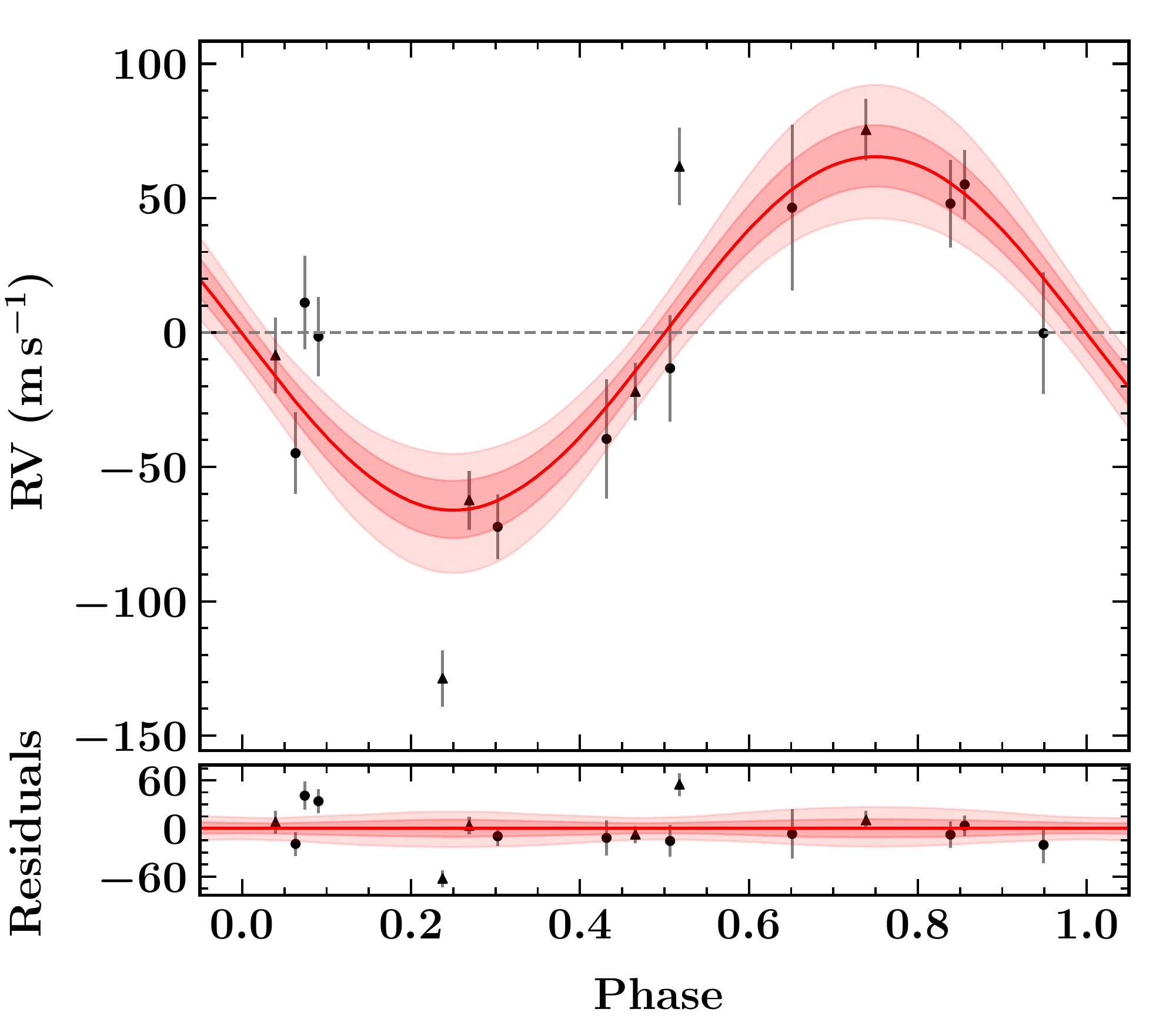}
    \caption{\textit{Top:} HARPS radial velocity curve of \Nstar\,, phase-folded to the best-fitting period as determined from the global modelling with GP-EBOP set out in Section~\ref{subsec:global_modelling}. Black points represent HARPS/HAM data points whereas black triangles show HARPS/EGGS mode data. The red line and pink shaded regions show the median and 1 \& 2\,$\sigma$ confidence intervals of the posterior model. \textit{Bottom:} Residuals of the fit with RMS $2\%$ and $3\%$ for HAM and EGGS respectively. Error bars and 1 and 2-sigma confidence regions are plotted as in the top panel.}
    \label{fig:rv}
\end{figure}

The RVs, along with associated properties, are detailed in Table~\ref{tab:rvs}. We plot the RVs, phase-folded on the best fitting period as determined by our global modelling analysis, in Figure \ref{fig:rv}.

\begin{table*}
	\centering
		\caption{Radial Velocities for \Nstar{}, acquired with HARPS on the ESO 3.6m telescope.}
	\label{tab:rvs}
	\begin{tabular}{lccccccc}
\multicolumn{1}{c}{BJD$_\mathrm{TDB}$}	&	RV		&RV error &	FWHM& 	Contrast & BIS & Exptime &Instrument\\
\multicolumn{1}{c}{(-2,450,000)}	& (\kms)& (\kms)&(\kms) & (\%)& (\kms)&(s)  & mode \\
		\hline
7959.563593&	-26.401&	0.022&	20.287&	12.1&	-0.0077& 1200 & HAM\\
7960.554740&	-26.315&	0.031&	20.336&	12.1&	-0.0768& 1200 & HAM\\
7962.536130&	-26.363&	0.015&	20.728&	12.0&	0.0879& 1200 & HAM\\
7979.521182&	-26.306&	0.013&	20.800&	11.9&	-0.0801 & 2400 & HAM\\
7980.507502&	-26.351&	0.017&	20.644&	12.0&	0.1010 & 1200 & HAM\\
7981.537916&	-26.434&	0.012&	20.307&	12.1&	0.2567 & 2400 & HAM\\
7997.489848&	-26.314&	0.016&	20.560&	12.1&	0.0628 & 2400 & HAM\\
7998.503038&	-26.407&	0.015&	20.566&	12.0&	0.0161 & 2400 & HAM\\
8000.503646&	-26.375&	0.020&	20.108&	12.2&	0.0935 & 2400 & HAM\\
8002.498343$^{*}$ &	-26.362&	0.023&	20.328&	12.0&	-0.2109 & 2400 & HAM\\
7982.509250&	-26.341&	0.014&	20.241&	12.3&	0.0500 & 1200 &	EGGS\\
7983.505570&	-26.328&	0.011&	20.100&	12.3&	-0.0542 & 1200 & EGGS\\
8160.797565& 	-26.412&	0.014&	21.057&	11.8&	-0.1349 & 1200 & EGGS\\
8161.832513&	-26.465&	0.011&	20.916&	11.9&	-0.0520 & 1200 &	EGGS\\
8188.757289&	-26.532&	0.011&	20.466&	12.1&	0.1690 & 1200&	EGGS\\
8189.787264&	-26.425&	0.011&	20.276&	12.2&	0.0519 & 1200&	EGGS\\
	\hline
    \end{tabular}
\begin{list}{}{\leftmargin=0em \itemindent=8.5em}
 \item[* Corrected for moonlight contamination using simultaneous sky fibre]
 \end{list}
\end{table*}

\section{Analysis}
\label{sec:analysis}

\subsection{Stellar properties}
\label{sec:stellar_properties}
\begin{table}
	\centering
	\caption{Stellar Properties for \Nstar.}
    \resizebox{\columnwidth}{!}{%
    \renewcommand{\arraystretch}{1.1}
	\begin{tabular}{lcc}
	Property	&	Value	&	Source\\
	\hline
    \multicolumn{3}{l}{Other Names}\\
    2MASS I.D.	&J14202949-3112074	& 2MASS \\
    %Gaia I.D. &DR1 6220602379784456320	& Gaia \\
    Gaia I.D. &DR2 6220602384081327104	& Gaia \\
    %NGTS I.D. &J142029.5-311206.7 & This work \\
    %NGTS object of interest I.D. & \NstarNOI & This work \\
    %NGTS field-object I.D. & \NstarLong{} & This work \\
    \\
    \multicolumn{3}{l}{Astrometric Properties}\\
    R.A.		&	14:20:29.5		&2MASS	\\
	Dec			&	-31:12:06.68	&2MASS	\\
    $\mu_{{\rm R.A.}}$ (\masy) & -21.736$\pm$0.093 & Gaia \\
	$\mu_{{\rm Dec.}}$ (\masy) & -0.858$\pm$0.090 & Gaia\\
    Parallax  (mas) & $2.779\pm0.063$& Gaia\\
    Distance  (pc) & \Ndist & SED fitting\\
    \\
   
    \multicolumn{3}{l}{Photometric Properties}\\
	V (mag)		&10.961$\pm$0.011 	&APASS\\
	B (mag)		&11.410$\pm$0.02	&APASS\\
	g (mag)		&11.121$\pm$0.008	&APASS\\
	r (mag)		&10.878				&APASS\\
	i (mag)		&10.771$\pm$0.019	&APASS\\
    G (mag)		&10.860				&{Gaia}\\
    NGTS (mag)	&10.790				&This work\\
    J (mag)		&10.055$\pm$0.023	&2MASS\\
   	H (mag)		&9.858$\pm$0.024	&2MASS\\
	K$_{\rm s}$ (mag) &9.799$\pm$0.021	&2MASS\\
    W1 (mag)	&9.748$\pm$0.023	&WISE\\
    W2 (mag)	&9.768$\pm$0.021	&WISE\\
    W3 (mag) &9.718$\pm$0.037 & WISE\\
    A$_{V}$	& $0.12\pm0.07$ & SED fitting\\
    L$_{0.1-2.4\,keV}$ (\luminosity) & $\leq\num{3.74e30}$ & This work \\
    L$_{2.0-12.0\,keV}$ (\luminosity) & $\leq\num{4.49e31}$ & This work \\
    L$_{\rm bol}$ (\luminosity) & $\num{1.76\pm0.13} \times 10^{34}$ & SED fitting \\
    Spectral type & F5V & SED fitting\\
    \\
    
    \multicolumn{3}{l}{Bulk Properties}\\
    \teff\,(K)    & \NstarTeff    &SED fitting\\
    log g (\cmss)& \Nstarlogg & Global modelling\\
    $\rho$ (\gcmthree)& \Nstardense & Global modelling\\
    $\left[Fe/H\right]$  & \FeH &HARPS spectra\\
	{\it v}\,sin\,{\it i} (\kms)	& \VSini	& HARPS spectra\\
	Age	(Gyrs) & \age	&	HARPS spectra\\
    Mass ($M_{\odot}$) &  \NstarMass & Global modelling\\
    Radius ($R_{\odot}$) &	\NstarRad & SED fitting\\
	\hline
    \multicolumn{3}{l}{2MASS \citep{2MASS}; APASS \citep{APASS}}\\
    \multicolumn{3}{l}{WISE \citep{WISE}; Gaia \citep{Gaia2018}}\\
	\end{tabular}
    }
    \label{tab:stellar}
\end{table}

\subsubsection{Bulk properties}
\label{subsubsec:stellar_bulk_properties}
To determine stellar bulk properties for \Nstar{} we compared 3 different methods, which we set out in this section. We determined initial stellar parameters by co-adding our individual HARPS (HAM) spectra and measuring equivalent widths, following a similar method to \citet{Doyle2013}. Hereafter we refer to this method as Method 1. We obtained the following results: \mbox{$T_{\rm eff}=6500\pm100$~K,} \mbox{log g = $4.0 \pm 0.2$,} \mbox{[Fe/H] = $-0.06\pm0.09$} and \mbox{{\it v}\,sin\,{\it i} = $15.2\pm0.8$ \kms{}}. For \mbox{{\it v}\,sin\,{\it i}} we assumed a macroturbulent velocity \mbox{${\it v_{mac}} = 6.8\pm0.7$ \kms}~ based on the asteroseismic calibration of \citet{Doyle2014}. No lithium was seen in the spectrum, suggesting \Nstar{} is not a young star.

It is well known that accurate determination of the mass and radius of a transiting exoplanet crucially depends on the accuracy with which one can determine the mass and radius of the stellar host. Historically, stellar host properties for most exoplanets characterised in the literature have been estimated using theoretical models or empirically calibrated relations, as opposed to direct observables, as they have generally provided the greatest accuracy. Using these methods, planetary masses and radii can typically be calculated with uncertainties of 6\% and 5\%, respectively \citep{Stassun2017}.

To estimate the mass, radius and age of \Nstar{}, as well as to check our other parameters are consistent, we also analysed our co-added HARPS spectra with the SPECIES code developed by \citet{soto2018}. Hereafter we refer to this method as Method 2.  As in Method 1, SPECIES also uses the measurement of equivalent widths and applies local thermodynamic equilibrium, along with ATLAS9 model atmospheres \citep{Castelli2004}, to obtain the: temperature, metallicity, surface gravity and micro turbulence of the stellar photosphere. Rotational velocity is found by absorption line fitting with synthetic spectra; the macroturbulent velocity was estimated using the relation from \citet{dosSantos2016}. Mass, radius and age are obtained using the \texttt{isochrones}
 package \citep{Morton2015}, which interpolates through a grid of MIST isochrones \citep{Dotter2016}. Photometric properties and parallax from Table~\ref{tab:stellar} are also included as priors in the isochrone interpolation. We obtained the following results for \Nstar{}: \mbox{\teff = $6604\pm134$ K,} \mbox{log g = $4.16^{+0.11}_{-0.09}$ dex,} \mbox{[Fe/H] = $-0.11\pm0.13$,} \mbox{$M_*=1.32^{+0.09}_{-0.08}\, M_{\odot}$,} \mbox{$R_*=1.58\pm0.22\, R_{\odot}$,} \mbox{{\it v}\,sin\,{\it i} = $13.5\pm1.3$ \kms{}} (assuming \mbox{${\it v_{mac}} = 7.6\pm1.3$ \kms)} and \mbox{age = 2.17$\pm0.37$ Gyr}.

With the second release of Gaia mission data (DR2), five-parameter astrometric solutions are now available for 1.3 billion sources, with an uncertainty in parallax measurements of up to 0.04 milliarcsecond for G<15 \citep{Gaia2018}. For \Nstar{} and other Gaia sources, the most accurate determinations of stellar masses and radii can now be made from direct observables: bolometric flux, effective temperature and parallax; the main uncertainties stem from $T_{\rm eff}$ and $A_v$. Using this method, \citet{Stassun2017} reported expected uncertainties in planetary masses and radii of 5\% and 3\% respectively. Early revisions to the properties of known exoplanets and their host stars, by exploiting this increased accuracy, are starting to appear in the literature. \citet{Berger2018} find a systematic upscaling of planetary radii in the range 1-5 $R_{\oplus}$ and confirm a gap in the radius distribution of small, close-in planets around 2 $R_{\oplus}$ with incident fluxes > 200~$F_{\odot}$.

\begin{figure}
	\includegraphics[width=\columnwidth]{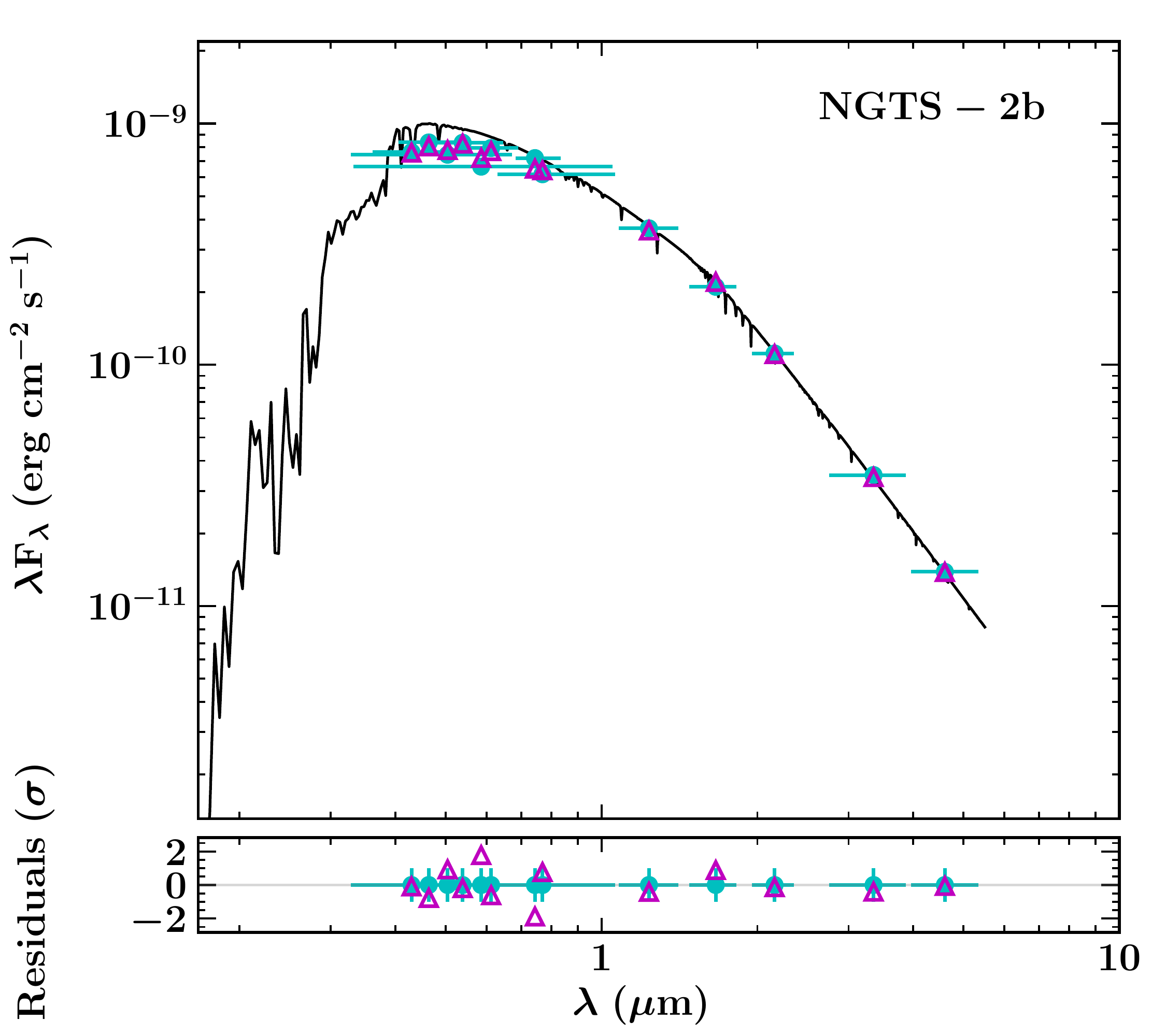}
    \caption{The fitted spectral energy distribution (black line) for \Nstar\ based on the photometric data (cyan points) presented in Table~\ref{tab:stellar} using the method presented in \citet{gillen17}. The black line shows the best fit model atmosphere and the magenta triangles show the model flux at the wavelengths of the photometric data. {\it Bottom}: Residuals of the fit in units of observational uncertainty.}
    \label{fig:sed}
\end{figure}

Utilising direct observables, we modelled the spectral energy distribution (SED) of \Nstar\ (Figure~\ref{fig:sed}), using the broadband photometric method described in \citet{gillen17}. Hereafter we refer to this method as Method 3. We modelled the SED by convolving PHOENIX v2 model atmospheres with the available bandpasses (see Table~\ref{tab:stellar}) and explored the posterior parameter space using Markov chain Monte Carlo (MCMC).
The parameters of the fit were the stellar temperature ($T_{\rm eff}$), surface gravity ($\log g$) and radius ($R$), the distance to the system ($d$) and the reddening along the line of sight ($A_{V}$). We also allowed a white noise jitter term ($\ln\sigma$) to account for additional uncertainties above the literature values. 
For $T_{\rm eff}$ and $\log g$, we used the values from the spectral modelling approach following \citet{Doyle2013} as priors in our fit. The radius, reddening and jitter terms had uninformative priors, and we constrained the distance using the Gaia DR2 parallax value \citep{Bailer-Jones15}. We obtain the following results: \mbox{$T_{\rm eff}$=\NstarTeff~K,} \mbox{$R_{*}$=\NstarRad~$R_{\odot}$,} \mbox{$d$=\Ndist~pc} and \mbox{$A_{V}=0.12\pm0.07$}. We note that our distance estimate is consistent with the Gaia-derived value presented in \citet{Bailer-Jones18}. See \citet[][and {\it in prep.}]{gillen17} for further details of the SED modelling protocol. 

In summary, we have calculated stellar parameters from 3 different methods:
\begin{enumerate}[(1)]
  \item $T_{\rm eff}$, log g, [Fe/H] and {\it v}\,sin\,{\it i} - from spectral-based method following \citet{Doyle2013}, applied to HARPS spectra
  \item $T_{\rm eff}$, log g, [Fe/H], {\it v}\,sin\,{\it i}, $M_{*}$, $R_{*}$ and age  - from the spectral-based method of \citet{soto2018} applied to HARPS spectra. Mass, radius and age were determined by interpolating in a grid of isochrones and Gaia parallax and broadband photometry were included as priors
  \item $T_{\rm eff}$, log g, $R_{*}$ and distance - from the broadband photometric SED fitting method of \citet{gillen17}. Gaia DR2 parallax was utilised along with priors on $T_{eff}$ and \mbox{log g} from Method 1.
\end{enumerate}

Comparing our results for \Nstar{} derived using the three different methods, we find that they are consistent within errors. We note that Method 3 (SED fitting) is the most data-driven approach with less dependence on models. Therefore in the remainder of this paper we adopt parameter values according to the following order of precedence:
\begin{enumerate}
 \item method 3
 \item method 1
 \item method 2
\end{enumerate}

Specifically, for $T_{\rm eff}$ and $R_{*}$ - we adopt values from Method 3; for [Fe/H] and vsini - we adopt values from Method 1; only Method 2 produced a value for age.

We have determined initial stellar parameters independently of the planet/system. In Section \ref{subsec:global_modelling} we subsequently consider the system as a whole to derive additional parameters. We apply global (simultaneous) modelling of all the data sets, incorporating the adopted values derived in this section as priors. We derive the stellar density from the best fitting transit model parameters, and combine the density with the stellar radius to calculate the stellar mass ($M_{*}$) and to further constrain \mbox{log g}. The final stellar parameters are summarised in Table~\ref{tab:stellar} with corresponding sources.

We investigated the X-ray brightness of the host star and found that no detection has yet been made at the source position. Nevertheless, using ROSAT and XMM-Newton slew images we were able to determine upper limits for the luminosity of $\sim\num{3.73e30}$ \luminosity{} in the 0.1--2.4 keV band and $\sim\num{4.47e31}$ \luminosity{} in the 2.0--12.0 keV band. As such, we conclude that \Nstar{} is fainter than a small percentage of the brighter F stars seen but cannot rule out that \Nstar{} itself is still among the brightest \citep{Panzera1999}.

\subsubsection{Kinematics}
Considering the proper motion (R.A.:\NstarPMRA\,\masy, Dec.: \NstarPMDec\,\masy), the absolute radial velocity (\Ngamma\,\kms), and the parallax distance of \Ndist\,pc, we calculate the ($U_{LSR},V_{LSR},W_{LSR}$) galactic motion for \Nstar{} to be (\GalULSR, \GalVLSR, \GalWLSR)\,\kms.  Given that these velocity vector components are small, this suggests that \Nstar\ is a member of the thin disk population.

\subsection{Stellar Activity and Rotation}
\label{sub:stellar_activity_rotation}
Knowledge of the stellar rotation period and activity properties of exoplanet host stars are valuable for a number of reasons. On the one-hand, activity can act as a nuisance in the radial velocity follow-up of transiting planets, since plage and starspots can manifest themselves as apparent RV shifts that can mask or (in extreme cases) mimic orbiting planets \citep[e.g.][]{Queloz2001, Huelamo2008}. In addition, the presence of unocculted spots in transit lightcurves can also systematically bias the determined planetary radii \citep[e.g.][]{Desert2011, Sing2011}. Hence understanding stellar activity can help counter such issues, thereby improving the veracity of the measured planetary parameters such as mass and radius. However, magnetic activity may also reveal the stellar rotation period, enabling additional important properties of the star and the planetary system as a whole to be constrained. For example, this includes system aging via gyrochronology \citep[e.g.][]{Barnes2007, Barnes2016}, as well as permitting highly misaligned transiting planetary systems to be identified through determination of the inclination of the stellar spin axis without the need for Rossiter-McLaughlin observations \citep[e.g.][]{Watson2010, Simpson2010, Schlaufam2010}.

\subsubsection{Spectroscopic constraints}
While \Nstar{} is a moderately rapid rotator with a $v \sin i_*$ of \VSini{} \kms (and hence has the potential to be an active star), this possibility is offset by its relatively early spectral type (F5V). This places it close to the boundary where the dynamo generating tacholine may not operate efficiently due to an extremely shallow outer convective envelope. However, the stellar parameters (see Table~\ref{tab:stellar}) allow an estimate of the expected stellar rotation period ($P_{rot}$) independent of the presence of rotationally modulated activity via:

\begin{equation}
P_{rot} = \bigg( \frac{2 \pi R_*}{v \sin i_*} \bigg)\sin i_*
\label{eqn:prot}
\end{equation}

\noindent where $R_*$, $v$ and $i_*$ are the stellar radius, stellar rotational velocity and inclination of the stellar rotation axis, respectively. Assuming spin-orbit alignment ($i_*$ = 90$^{\circ}$), this results in $P_{rot}$ = \Prot days -- which also represents the upper limit to $P_{rot}$.

\subsubsection{Photometric constraints}
We performed a detailed search for the signal of a photometric rotation period in NGTS data. In this analysis, the transits of \Nplanet{} were removed (just leaving the out-of-transit light) prior to pre-whitening to also remove integer 1- and 2-day periods that occur due to the observing window function.

\begin{figure}
    \includegraphics[width=\columnwidth]{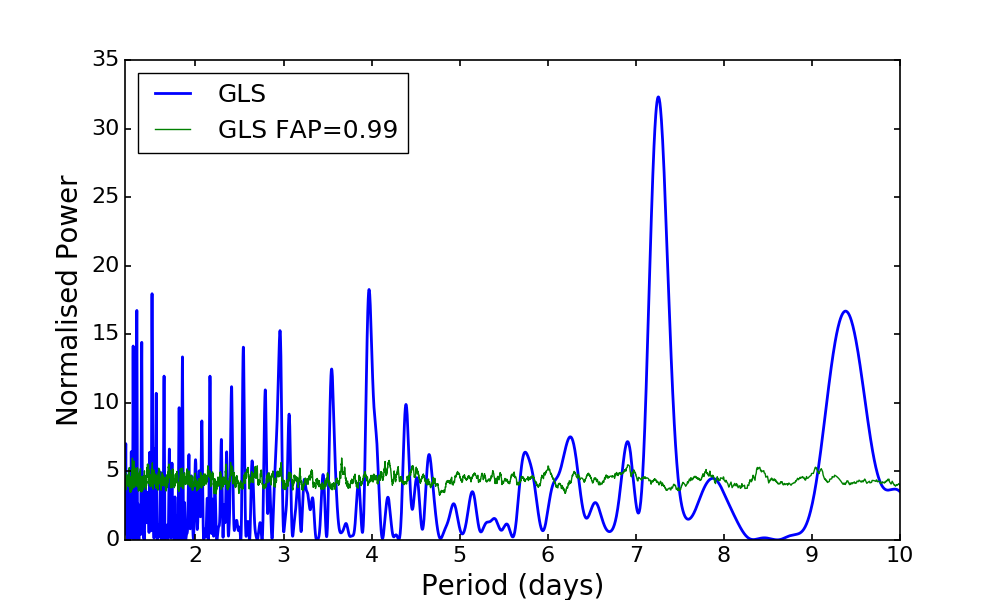}
    \caption{Generalised Lomb Scargle Periodogram for \Nstar{}, after detrending for the window function and removal of transit signals. There are strong signals at 7.25 and 3.97 days.}
    \label{fig:gls}
\end{figure}

Figure~\ref{fig:gls} shows the generalised Lomb Scargle (GLS) periodogram spanning 1.2--10 days (thought to encompass the likely rotation period of \Nstar{} -- see earlier), along with a false alarm probability (FAP) calculated using 1,000 bootstraps of the NGTS photometry in order to sample the window function in that period range. This shows a main peak at 7.25 days, with the next strongest peak lying at 3.97 days (as well as a forest of other signals, predominantly at lower periods). We carried out a wavelet analysis of the same data but binned into $\sim$15 minute intervals to allow for a constant time interval required for wavelet analysis \citep{Torrence1998}. Gaps in the data were left as is, and this does not have a considerable effect on the resultant power spectrum. 

Both the wavelet and GLS show signals at $\sim$7.2 and 3--4 days, although 7.2 days is not consistent with the estimated rotational period, which constitutes an upper limit of \Prot days. The wavelet analysis shows two distinct timespans where these signals are strongest (from $\sim$70 -- 120 days, and again from $\sim$150 -- 200 days). We have phase-folded the NGTS light-curves on the prominent periods for the entire dataspan, as well as only for times when the wavelet analysis indicated high power at those periods. We see no clear evidence of a rotationally modulated signal in any of these phase folded light curves.

A median periodogram was produced for the $\sim$11,000 other stars within the same \NGTS{} field as \Nstar{}, observed in the same season and with the same camera. This shows a very broad, minor peak covering periods between $\sim$6 and $\sim$8 days, significant periodicity around $\sim$4 days, as well as a forest of power at shorter periods.

We conclude that the observed peaks for \Nstar{} at $\sim$4 days and 7.2 days likely arise due to systematics. We see no convincing evidence for a rotational period of \Nstar{} in the \NGTS{} light curve, which is consistent with the estimated value irrespective of whether the system is aligned or misaligned.

\subsection{Global Modelling}
\label{subsec:global_modelling}
Intrinsic stellar variability (Section \ref{sub:stellar_activity_rotation}) and residual observational systematics can give rise to time correlated (``red'') noise in photometric light curves, which can subtly alter the shape of transits leading to inaccurate model fitting \citep{Pont2006,Pont2008,Silva-Valio2008}. 

Given the presence of correlated noise (likely stellar and systematic) in the NGTS light curve, we globally modelled \Nstar{} using \gpe\ \citep{gillen17} to constrain the stellar and planetary parameters. \gpe\ comprises a central EBOP-based transiting planet and eclipsing binary model, which is wrapped within a Gaussian process (GP) model. The GP is designed to robustly account for stellar activity and instrumental systematics, and propagate uncertainties due to these into the posterior distributions of the stellar and planetary parameters. \gpe\ explores the posterior parameter space using EMCEE \citep{Foreman-Mackey2013}. The interested reader is referred to \citet{gillen17} for further details of \gpe's modelling protocol. 

Before modelling with \gpe, the detrended NGTS light curve was normalised by the median out-of-transit flux and binned to 10 mins cadence. A quadratic limb darkening law was utilised (\citealt{Kopal1950,Mandel2002}) with the profiles and uncertainties obtained via the LDTk package \citep{Husser2013,Parviainen2015}. LDTk was given estimates of $T_{\rm eff}$, $\log g$ and [Fe/H] from our SED modelling and the spectral modelling method of \citet{Doyle2013}. The resultant uncertainty in the limb darkening profile was inflated by a factor of 10 to account for systematic errors in the stellar atmosphere models in the region of parameter space where \Nstar{} lies.

We fitted for an systemic radial velocity offset between HARPS/HAM and HARPS/EGGS modes. In addition to the offset, \gpe\ also allows for stellar and systematic RV jitter above the observational uncertainties, with data from each instrument treated individually. Given the relatively short orbital period of \Nplanet, theoretical and empirical evidence of single systems would favour a circular orbit \citep{Anderson2012}. Furthermore, the observational RV uncertainties likely preclude a robust detection of eccentricity below a moderate value. Nevertheless we compared a circular orbit model (e=0) with a model where eccentricity was free to vary.

In both cases, we stepped through the parameter space 50,000 times with each of 150 walkers (conservatively discarding the first 30,000 steps as burn in). Walkers were initialised from within a representative region of parameter space. Chains were thinned by a factor of 100 to reduce clustered samples, unrepresentative of the true posterior distribution, due to autocorrelation. Finally, the Gelman-Rubin criterion \citep{gelman1992} was used to check chain convergence.

Comparing the results from the eccentric and circular models, we reassuringly find consistent results for our parameters of interest. However, we note that the eccentricity is not well constrained ($e=0.035^{+0.106}_{-0.031}$). Given the lack of secondary eclipse, the main constraint on eccentricity is given by the RV data, which have large uncertainties ($\sim20$ \ms). We identified two families of low eccentricity solution, which are compatible with the data, differing in their eccentricity combination terms but not their overall eccentricity value. The eccentricity value and uncertainty should therefore be treated with caution and thus we adopt the circular model as our main model.

We find that \Nstar{} is orbited by \Nplanet{} with semi-major axis a=\SemiMajorAxis{}~AU and inclination i=\OrbitalInc{}~deg. \Nplanet{} has a mass \NplanetMass{}~\mjup, radius \NplanetRad~\rjup{} and density \NplanetDense~\gcmthree. Assuming an albedo equal to that of Jupiter, we calculate an equilibrium temperature for \Nplanet{} of \NplanetTeff~K but note that such assumption has a large uncertainty given the lack of knowledge of the planetary atmosphere \citep{Borucki2011}. Interestingly, we emphasise that this is a low planetary density for a HJ and discuss this in Section \ref{sec:discussion}. In agreement with previous estimates from spectral based methods, we determine a stellar mass and log g for \Nstar{} of \NstarMass~\msun{} and \Nstarlogg~\gcmthree{} respectively, by combining the adopted stellar radius from SED fitting with the stellar density measured directly from the posterior parameters. We adopt these values as the final mass and log g for \Nstar{} and include them in the summary table of stellar properties (Table \ref{tab:stellar}). Fitted and derived parameters from our global modelling are presented in Table \ref{tab:system_params}.

\begin{table*}
\centering
\caption{Planetary and system parameters from global modelling of the \Nstar{} system. The results from fixed eccentricity (e=0) versus floating eccentricity fitting are compared. The median values of the posterior distributions were adopted as the most probable parameters, with the $1\sigma$ intervals as the error estimates. We present results from the circular orbit (e=0) scenario as our main results (see Section \ref{subsec:global_modelling} for explanation).}
\label{tab:system_params}
\renewcommand{\arraystretch}{1.5}
\begin{tabular}{ccccc}
\hline
\multirow{2}{*}{\bfseries Parameter} &
\multirow{2}{*}{\bfseries Description} &
\multirow{2}{*}{\bfseries Unit} &
\multicolumn{2}{c}{\bfseries Value} \\
\cline{4-5}
& & &
{\bfseries Fixed e=0} &
{\bfseries Floating e} \\
\hline

Fitted parameters \\ 
\cmidrule(lr){1-5}
$\frac{R+r}{a}$ & Sum of radii relative to semi-major axis of system & none & $0.1369^{+0.0065}_{-0.0028}$ & $0.1377^{+0.0080}_{-0.0037}$
\\
$k$ & radius ratio planet to star, $\frac{r}{R}$ & none & $0.09619^{+0.00114}_{-0.00088}$ & $0.09633^{+0.00136}_{-0.00097}$
\\
$\cos i$ & cosine of orbital inclination & none & $0.026^{+0.021}_{-0.018}$ & $0.030^{+0.023}_{-0.021}$
\\
$\sqrt{e} \cos \omega$ & orbital eccentricity and argument of periastron term & none & & $0.00^{+0.10}_{-0.13}$ 
\\
$\sqrt{e} \sin \omega$ & orbital eccentricity and argument of periastron term & none &
& $0.050^{+0.093}_{-0.326}$ 
\\
$P$ & orbital period & days & $4.511164\pm0.000061$ & $4.511164^{+0.000070}_{-0.000069}$
\\
$T_{c}$ & epoch of transit centre & BJD & $2457759.1261^{+0.0014}_{-0.0013}$ & $2457759.1259^{+0.0016}_{-0.0017}$
\\
$q1_{\rm{NGTS}}$ & first Kipping LD term\,$^{*}$ & none & $0.3431^{+0.0072}_{-0.0071}$ & $0.3431^{+0.0073}_{-0.0072}$
\\
$q2_{\rm{NGTS}}$ & second Kipping LD term\,$^{*}$ & none & $0.3903^{+0.0055}_{-0.0054}$ & $0.3904\pm0.0055$
\\
$\ln(\sigma^{2})_{\rm{HAM}}$ & natural log of jitter in HAM RV data & ln(\kmssquared) & $-10.6^{+2.6}_{-6.3}$ & $-10.3^{+2.5}_{-6.3}$
\\
$\ln(\sigma^{2})_{\rm{EGGS}}$ & natural log of jitter in EGGS RV data & ln(\kmssquared) & $-6.56^{+0.85}_{-0.73}$ & $-6.52^{+0.83}_{-0.74}$
\\
{$K$} & radial velocity semi-amplitude of star & \ms & $65.8^{+9.5}_{-9.1}$ & $66.1^{+9.8}_{-9.2}$
\\
$V_{\rm{sys}}$ & systemic velocity & \kms & $-26.3616^{+0.0064}_{-0.0063}$ & $-26.3615^{+0.0063}_{-0.0064}$
\\
$\delta$V$_{\rm{sys~EGGS}}$ & EGGS RV  mode offset & \ms & $41^{+19}_{-18}$ & $41^{+18}_{-19}$
\\
$\ln(A^{2})_{\rm NGTS}$ & natural log of squared amplitude $\ddagger$ & ln(rel. flux$^2$) & $-13.90\pm0.13$ & $-13.89\pm0.13$
\\
$\ln(l^{2})_{\rm NGTS}$ & natural log of squared timescale $\ddagger$ & ln(days$^2$) & $-9.36^{+0.32}_{-0.30}$ & $-9.37^{+0.34}_{-0.31}$
\\
$\ln(\sigma^{2})_{\rm NGTS}$ & natural log of variance $\ddagger$ & ln(rel. flux$^2$) & $-29.7^{+6.9}_{-7.0}$ & $-29.8^{+6.8}_{-6.9}$
\\
\cmidrule(lr){1-5}
Derived parameters \\
\cmidrule(lr){1-5}
$e$ & orbital eccentricity & none & & $0.035^{+0.106}_{-0.031}$
\\
$i$ & orbital inclination & $deg$ & \OrbitalInc & $88.3^{+1.2}_{-1.3}$
\\
$a$ & semi-major axis of system & AU & \SemiMajorAxis & $0.0628^{+0.0026}_{-0.0035}$
\\
$r$ & radius of planet & $R_{jup}$ & \NplanetRad & $1.598^{+0.048}_{-0.046}$
\\
$m$ & mass of planet & $M_{jup}$ & \NplanetMass & $0.76^{+0.17}_{-0.15}$
\\
$\rho$ & density of planet & \gcmthree & \NplanetDense & $0.232^{+0.054}_{-0.048}$
\\
$T_{eq}$ & equilibrium temperature of planet & $K$ & \NplanetTeff & $1472^{+50}_{-44}$
\\
$T_{14}$ & transit duration & hours & \NplanetDuration & $4.67^{+0.38}_{-0.11}$
\\
\hline
\end{tabular}
\begin{list}{}{}  
 \item[* LD = limb darkening; see \citet{Kipping2013} for a detailed description]
 \item[$\ddagger$ Gaussian Process Hyperparameters]
 \end{list} 
\end{table*}

\section{Discussion and conclusion}
\label{sec:discussion}
\begin{figure}
	\includegraphics[width=\columnwidth]{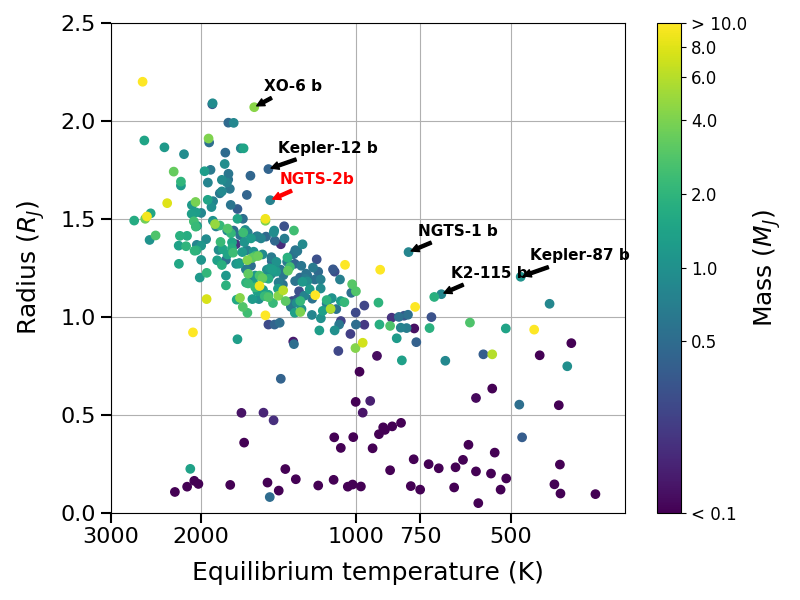}
    \caption{Parameter space of planetary radius versus planetary equilibrium temperature, for confirmed exoplanets\protect\footnote{}. Planetary mass is indicated by the colour-bar scale. The location of \Nstar{} is shown by a red arrow, assuming an albedo equal to that of Jupiter, while a selection of inflated planets are shown by black arrows. The combination of lower temperature, lower mass but higher radius of \Nplanet{}, compared to the distribution of planets, highlights that \Nplanet{} is inflated.}
    \label{fig:planet_param_space}
\end{figure}

\begin{figure}
	\includegraphics[width=\columnwidth]{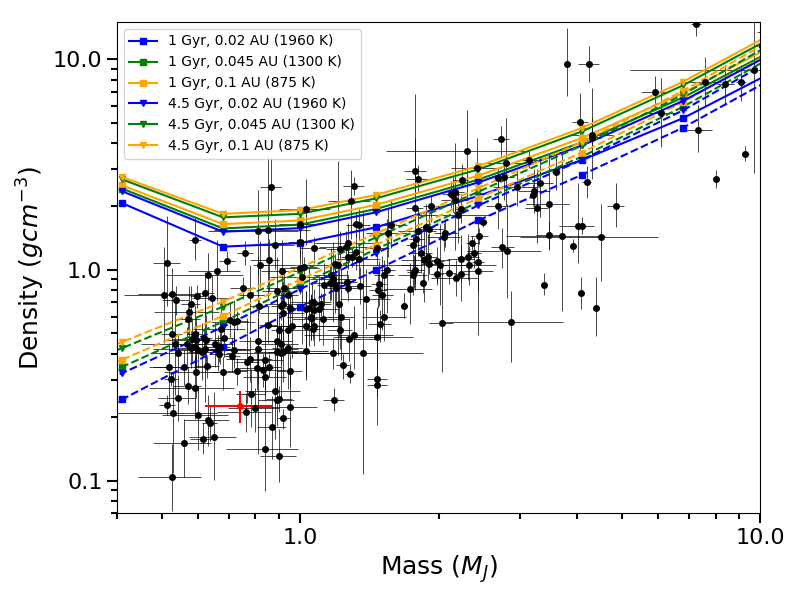}
    \caption{Planetary density versus planetary mass, for previously confirmed HJs$^{1}$ (black data points) and \Nplanet{} (red data point). Theoretical relations from \citet{Fortney2007} are plotted for comparison, considering planetary evolution under stellar irradiation. We plot theoretical relations for planet ages, orbital separations and effective temperatures similar to \Nplanet. Dashed lines depict planets with $100~M_{\oplus}$ cores, where as solid lines represent planets with no cores. The density of \Nplanet{} exceeds the theoretical values for a planet of the same age and effective temperature, irrespective of its composition. This suggests additional heating mechanisms, which are not considered in the theoretical models, are contributing to the atmospheric heating of \Nplanet.}
    \label{fig:planet_radius_mass_models}
\end{figure}

With a radius of \NplanetRad~R$_{\rm J}$, mass \NplanetMass~\mjup{} and equilibrium temperature \NplanetTeff~K, \Nplanet{} is an inflated HJ (Figure~\ref{fig:planet_param_space} \& \ref{fig:planet_radius_mass_models}). In fact the density of \Nplanet{} is only \NplanetDense~\gcmthree{}, placing it among the least dense planets known. Orbiting with period \NperiodShort{} days, \Nplanet{} is slightly further from its stellar host than most HJs with similar densities and host spectral types. For instance, WASP-121b \citep{Delrez2016} orbits its F6V star much closer in, with period 1.27 days.

Other HJs with masses and radii similar to \Nplanet{} include WASP-90b \citep{West2016}, WASP-118b \citep{Hay2016,Mocnik2017} and WASP-88b \citep{Delrez2014}. The lower density of these planets suggests a planetary heating mechanism is at play, which is independent of stellar irradiation. Plausible explanations could include increased internal heat generation \citep{Ginzburg2015}, double-diffusive convection \citep{Leconte2012} and increased atmospheric opacity \citep{Burrows2007}. In order to properly ascertain the causes of HJ inflation, the sample of inflated planets with precisely measured masses and radii must be expanded, allowing trends to be discerned which distinguish the competing theories.

Characterisation of exoplanet atmospheres via transmission spectroscopy has been carried out for bright targets using HARPS \citep{Wyttenbach2015}, ESPRESSO \citep{pepe2010} and  HST \citep{sing2016}. The \JWST, scheduled for launch in 2020, will further enable atmospheric measurements at a higher level of precision \citep{Beichman2014,Stevenson2016}. \Nplanet{} is a short-period gas giant orbiting a bright (V$\sim$11) stellar host. In addition, we calculate an atmospheric scale height of $\sim760$ km. These properties make \Nplanet{} an ideal target for such studies. 

Previous work has shown that the orbits of hot and warm Jupiters around early-type stars are often misaligned with the stellar rotation axis \citep[e.g.,][]{winn15}. The effective temperature of NGTS-2 (\NstarTeff\,K) is hotter than the empirical boundary of $\simeq$6250\,K, above which high orbital obliquities are common (particularly for giant planets with $M \lesssim 3$M$_{\mathrm J}$).  
We were not able to determine a convincing rotational period of \Nstar{} from our photometric and spectroscopic analysis  (Section \ref{sub:stellar_activity_rotation}), accordingly we cannot confirm whether there is spin-orbit misalignment in the \Nstar{} system. We recommend Rossiter-McLaughlin follow-up observations and note that either outcome (aligned or misaligned) is interesting for the stellar type of \Nstar. The brightness of the host star combined with the rapid stellar rotation should make the effect readily detectable. Using Eq. (40) from \citet{Winn2010} we calculate a maximum amplitude for the signal of $110$~\ms, which is an order of magnitude larger than our typical HARPS RV measurement error of $\sim20$~\ms.

The Transiting Exoplanet Survey Satellite (\TESS; \citealt{Ricker2014}) has recently launched, with the Guest Investigator Program allowing observations to be made outside of the mission's core science operations. We searched for, and found no evidence of, multiple planets in the \Nstar{} system. As a space-borne observatory, TESS' increased photometric precision over ground-based facilities is more equipped to search for additional (and shallower) transit signals, indicative of other planets in this system besides \Nplanet. The presence of multiple planets can also be inferred from Transit Timing Variations (TTVs), as was achieved with K2 observations of WASP-47 \citep{Becker2015}. The 2 minute cadence would be required for robust TTVs.

% Reset footnote counter
\setcounter{footnote}{1}

\footnotetext{\url{https://exoplanetarchive.ipac.caltech.edu}, \mbox{online 16 March 2018}} 

In conclusion we have discovered \Nplanet{}, an inflated hot-Jupiter (\mbox{M$_p$=\NplanetMass~\mjup,} \mbox{R$_p$=\NplanetRad~\rjup,} \mbox{$\rho_p$=\NplanetDense{}~\gcmthree)} transiting a bright F5V star in a \NperiodShort{} day orbit. \Nplanet{} is one of the least dense exoplanets currently known. Selection of this ideal target for future follow-up studies may advance our understanding of the formation and evolution of hot-Jupiters and their atmospheres.

The considerable power of \NGTS{} as an exoplanet survey facility, has been demonstrated by confirming \Nplanet{} without the need for follow-up photometry. Over a 4 year period, \NGTS{} is expected to yield $\sim$ 200 planets larger than Neptune and more importantly $\sim$10 smaller planets \citep{Guenther17a}, all orbiting bright hosts. In the era of \TESS{}, \NGTS{} will undoubtedly play a crucial role in candidate follow-up. The enhanced plate scale of \NGTS{} (5\,\arcsecpix) compared to \TESS{} (21\,\arcsecpix) will allow better separation of blended targets in the \TESS{} fields. In addition, based on noise models for a 1 hour sampling rate (\citealt{Ricker2014,Wheatley2018}) \NGTS{} is expected to achieve higher photometric precision than \TESS{} for magnitudes fainter than $I=14$, owing to its larger aperture.  

\section*{Acknowledgements}
% NGTS at ESO
This publication is based on data collected under the NGTS project at the ESO La Silla Paranal Observatory. The NGTS instrument and operations are funded by the consortium institutes and by the UK Science and Technology Facilities Council (STFC; project reference ST/M001962/1). 
% Leicester
LR is supported by an STFC studentship (1795021). The contributions at the University of Leicester by MRG and MRB have been supported by STFC through consolidated grant ST/N000757/1. SLC acknowledges support from LISEO at the University of Leicester.
% RDA at Leicester
This project has received funding from the European Research Council (ERC) under the European Union's Horizon 2020 research and innovation programme (grant agreement No 681601).
% Warwick
The contributions at the University of Warwick by PJW, RGW, DJA, DP and TL are supported by an STFC consolidated grant (ST/P000495/1).
% Chile
JSJ acknowledges support by FONDECYT grant 1161218 and partial support by CATA-Basal (PB06, CONICYT).
% Cambridge
MNG is supported by the UK STFC award reference 1490409 as well as the Isaac Newton Studentship.
% Geneva
Contributions at the University of Geneva by FB, BC, LM, and SU were carried out within the framework of the National Centre for Competence in Research "PlanetS" supported by the Swiss National Science Foundation (SNSF).
% Queens
CAW acknowledges support from the STFC grant ST/P000312/1.
% Gaia
This work has made use of data from the European Space Agency (ESA) mission
{\it Gaia} (\url{https://www.cosmos.esa.int/gaia}), processed by the {\it Gaia}
Data Processing and Analysis Consortium (DPAC,
\url{https://www.cosmos.esa.int/web/gaia/dpac/consortium}). Funding for the DPAC
has been provided by national institutions, in particular the institutions
participating in the {\it Gaia} Multilateral Agreement.

%%%%%%%%%%%%%%%%%%%%%%%%%%%%%%%%%%%%%%%%%%%%%%%%%%

%%%%%%%%%%%%%%%%%%%% REFERENCES %%%%%%%%%%%%%%%%%%

% The best way to enter references is to use BibTeX:

\bibliographystyle{mnras}
\bibliography{paper}

%%%%%%%%%%%%%%%%%%%% AFFILIATIONS %%%%%%%%%%%%%%%%%%
\section*{Affiliations}
{\it
$^{l}$Department of Physics and Astronomy, Leicester Institute of Space and Earth Observation, University of Leicester, LE1 7RH, UK\\
$^{c}$Cavendish Laboratory, J.J. Thomson Avenue, Cambridge CB3 0HE, UK\\
$^{d}$Institute of Planetary Research, German Aerospace Center, Rutherfordstrasse 2, 12489 Berlin, Germany\\
$^{w}$Dept.\ of Physics, University of Warwick, Gibbet Hill Road, Coventry CV4 7AL, UK\\
$^{ce}$Centre for Exoplanets and Habitability, University of Warwick, Gibbet Hill Road, Coventry CV4 7AL, UK\\
$^{g}$Observatoire de Gen{\`e}ve, Universit{\'e} de Gen{\`e}ve, 51 Ch. des Maillettes, 1290 Sauverny, Switzerland\\
$^{q}$Astrophysics Research Centre, School of Mathematics and Physics, Queen's University Belfast, BT7 1NN Belfast, UK\\
$^{ca}$Institute of Astronomy, University of Cambridge, Madingley Rise, Cambridge CB3 0HA, UK\\
$^{uc}$Departamento de Astronomia, Universidad de Chile, Casilla 36-D, Santiago, Chile\\
$^{ci}$ Centro de Astrof\'isica y Tecnolog\'ias Afines (CATA), Casilla 36-D, Santiago, Chile.\\
$^{a}$Instituto de Astronomia, Universidad Cat\'{o}lica del Norte, Casa Central, Angamos 0610, Antofagasta, Chile\\
$^{fu}$Institute of Geological Sciences, FU Berlin, Malteserstr. 74-100, D-12249 Berlin, Germany\\
$^{tu}$Center for Astronomy and Astrophysics, TU Berlin, Hardenbergstr. 36, D-10623 Berlin, Germany\\
$^{k}$Astrophysics Group, Lennard-Jones Laboratories, Keele University, Staffordshire ST5 5BG, UK\\
$^{rs}$Dr. Remeis-Sternwarte, Friedrich Alexander Universit\"at Erlangen-N\"urnberg, Sternwartstr. 7, 96049 Bamberg, Germany\\
$^{\dagger}$\,Winton Fellow
}

%%%%%%%%%%%%%%%%%%%%%%%%%%%%%%%%%%%%%%%%%%%%%%%%%%

%%%%%%%%%%%%%%%%% APPENDICES %%%%%%%%%%%%%%%%%%%%%
%\appendix
%\section{Supplementary figures}
%\FloatBarrier
%%%%%%%%%%%%%%%%%%%%%%%%%%%%%%%%%%%%%%%%%%%%%%%%%%

% Don't change these lines
\bsp	% typesetting comment
\label{lastpage}
\end{document}